# Mathematical and computer tools of discrete dynamic modeling and analysis of complex systems in control loop

Armen G. Bagdasaryan

*Abstract*—In this paper we present a method of discrete modeling and analysis of multilevel dynamics of complex large-scale hierarchical dynamic systems subject to external dynamic control mechanism. In a model each state describes parallel dynamics and simultaneous trends of changes in system parameters. The essence of the approach is in analysis of system state dynamics while it is in the control loop. Architectural model of information system supporting simulation and analysis of dynamic processes and development scenarios (strategies) of complex large-scale hierarchical systems is also proposed.

*Keywords*—Discrete modeling, dynamics, complex systems, control, hierarchy, large-scale objects, computer system model.

## I. Introduction

Complex systems are all the real world systems that surround us. For example, social and economic systems, neural networks, artificial intelligence, computations, swarm of software agents, ecology, culture, traffic patterns, terrorist networks, biological systems, and many other scientific areas can be considered to fall into the realm of complex systems.

Complex systems contain a large number of mutually interacting entities (components, agents, processes, etc.) whose aggregate activity is not derivable from the summations of the activity of individual entities, and typically exhibit hierarchical self-organization. Another important characteristic of complex systems is that their description requires the notion of purpose, since the systems are generally purposive. Elements of complex system have their individual purposes. Achievement of these purposes contributes to the corporate system purpose but at the same time purposes of elements and purpose of the whole system are as a rule in a conflict.

Any scientific method (approach) of studying complex real world systems relies on modeling (analytical, numerical) and computer simulation. Among the analytical techniques are statistical mechanics, stochastic dynamics, non-equilibrium thermodynamics, etc. Among the computer simulation techniques are cellular automata, multi-agent techniques, evolutionary programming, Monte Carlo methods, etc. Since analytical treatments alone do not allow us to understand a complex system, computer simulations play a key role in our understanding of how these systems function and work. This is also true and possibly in a more degree for complex control systems. The main characteristic of modern complex control systems is that it is impossible to uniquely and adequately describe these systems, using classical mathematical methods. Classical mathematical models are suitable just for a few problem domains, which are static and comprehensible, and have most general properties. And there still remains a wide range of complex problems that can not be described by the existing formal methods.

Today we can distinguish several basic forms of complexity: structural (geometrical, topological), dynamical, hierarchical, algorithmic, and large scale. Taking into account the interplay between intellectualized mathematical and information technologies of control and decision support play an important role in modeling of processes of evolution and functioning of complex (large-scale) systems.

Complex systems are usually difficult to model, design, and control. In studying complex systems, the behavior of which depends on the elements interactions, an integrative system-theoretic (top-down) approach is more preferable, as compared to a reductionist (bottom-up) one. However, a compromise between both approaches should be found.

A central goal of this work is to propose models and modeling technique that are useful when applied to the complex systems, which can with a sufficient accuracy be described by models of development of hierarchical systems. Aiming at this, we develop method for constructing discrete models of complex hierarchical dynamic systems subject to external hierarchical dynamic control mechanism and their problem-oriented interpretation. The method includes:
1. Creating multivariate multilevel hierarchical structural model based on system analysis;
2. General mathematical formalization;
3. Constructing hierarchical dynamic graph model to solve a system development control problems and

to analyze system dynamic characteristics related to the attainability of desirable states and goals;
4. Specializing the model to the scenario-type schemes of control of complex systems.

## II. PROBLEM FORMULATION, GENERAL FEATURES AND DESCRIPTION OF MODEL

Modeling and analysis of control and dynamic processes in complex multi-component large-scale systems make it necessary to operate with multiple state coordinates. This is caused by: (1) the fact that complex large-scale system behavior is influenced by a number of factors of various nature which leads to large amount of system parameters, indicators, and variables; (2) lack of sufficient information (incompleteness, uncertainty) on the state and processes that influence system development, especially, for systems and objects belonging to weakly-formalizable ones. Therefore, a peculiar approach which will allow taking into account all essential diverse factors that determine system activity and behavior under the influence of external control actions is needed. The modeling technique developed allows one to cope with the above mentioned problems. The control and problem domains have the following features:

1. Multilevel dynamical systems consisting of a set of autonomous elements (subsystems) with local (individual) and global (corporate, general) problems and goals is considered as a canonical model of control object;
2. The external dynamic control mechanism in a system is considered as a set of control actions initiating multilevel state dynamics of control object;
3. The long-term databases and monitoring that characterize the changing of parameters and indicators can be used as the main source of information about system behavior, development and control problems. Databases, monitoring data and other statistical material facilitate observing for the changes in parameters at different time intervals, which has an extreme promise for understanding the global regularities in system dynamics. Monitoring includes observation of the current situation around the system. The processes under monitoring are interpreted in the form of state dynamics and estimations, and tendencies of system development as well. The system goals are formulated as consistent dynamics of these processes. Monitoring of the present situation enables (1) discovering new factors and parameter estimations influencing the system development, (b) establishing possible new or desirable states and goals. In this case the model is updated;
4. Due to the hierarchical structure of a parameter set, a multilevel (hierarchical) control loop based on a set of independent closed control loops of lower levels is constructed. The basic criteria for multilevel control loop efficiency are consistency and time-event coordination of attainable states (goals) and of dynamical properties of system parameters.

In our approach a large-scale hierarchical system is understood as a combination of distributed in time and space interacting subsystems that organize separate hierarchical levels. On each level a subsystem is assumed to be described in corresponding space of parameters and variables, some of them are so called polymorphic that equally applicable for objects at different levels of hierarchy. On each hierarchical level the system has its local goals.

A control problem of system development is considered as a construction of controlling scenario realizing a time-event coordination of control actions to achieve control goals of subsystems at different hierarchical levels and at the same time to implement global system goals.

In a whole, system functioning efficiency depends not only on the "top-down" influence but on the "bottom-up" response as well, i.e. on the consistent behavior of all system elements.

In connection with what has been said above the key requirements to the model of system development and control are the following:

1. Representation of interrelations between local and global goals;
2. Consistency and efficiency of control actions in the process of goals achievement (at the level of required values and dynamics of parameters);
3. Time and event ordering of needed control actions;
4. Problem-oriented significance of each control action.

The basic complexity of the problem under consideration is:

1. to make consistent and to coordinate a set of problems and conflicting goals in long-term system development;
2. to analyze tendencies, shifts, and value proportions in parameter changes, and classification of control objects;
3. to make consistent practically unlimited number of dynamical processes.

This raises the problem of choosing the principles which to a maximal extent reduce all the variety of dynamical interrelations when modeling a large-scale hierarchical system to clear and logic constructions. The following basic principles of system modeling have been chosen:

1. Systematic and necessary use of hierarchy (fig. 1). In our case, the hierarchy is used for representation of control domain and its qualitative characteristics (polymorphic parameters). The principle of hierarchy allows us to distinguish the essential interrelations for aggregation and scaling (recount) of dependent parameters, and also helps structuring the problem and control domain of an object;
2. Use of the concept of state as system indicators. The notion of state is used as a mean for description of combinability of values of various parameters, logically coupled and uncoupled;
3. Use of state diagrams in control loop. The efficiency of control actions and comparison for efficiency of different sets of control actions are formulated on the basis of state diagrams.

To cope with the problem of dimension, which is due to the growth of the model, the states are considered not as a combination of parameter values or its ranges but as a combination of parallel trajectories of parameter changes (principle of getting the integrated concepts of pseudo-organisational system). The description of trajectories includes not only primitive growth/drop of parameter values but also some typical paths of changes of a parameter value ranges, with stable and unstable cycles.

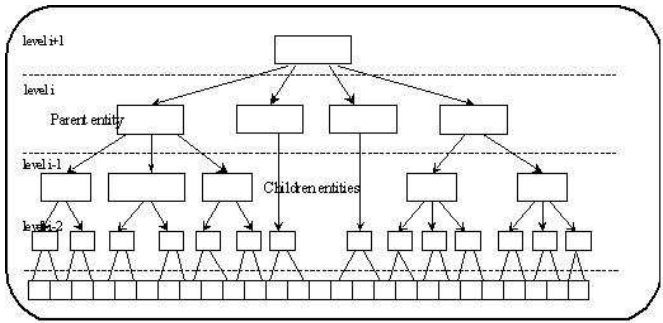

Fig. 1. Structuring control and problem domain

We include in one state a set of static state characteristics on time interval and get aggregated characteristics for practically unlimited number of parameters simultaneously. In combination with the principle of hierarchy this is a convenient tool for description of interrelations between tendencies, shifts and proportions in changes of parameter values related to the objects at different levels of system. This, in turn, serves as a basis of analysis for coordination of dynamics of various hierarchical parameters of system elements. The principles of modeling given above enable us to formulate a universal model of complex large-scale hierarchical dynamic system in control loop.

III. STATE DIAGRAMS AS A FORMALISM FOR REPRESENTATION OF SYSTEM DEVELOPMENT DYNAMICS

In this section we describe a model of development of complex hierarchical objects. We introduce hierarchical state diagrams as a tool for formal representation of system parameter dynamics. The dynamics of objects' parameter changes is the main body of the model.

*A. General Outline*

The system approach to the modeling of complex systems supposes the analysis of interconnected processes for as many components as it possible. To satisfy this requirement we organize the modeling so that the objects development is manifested in the form of state dynamics which characterizes both the system as a whole and its components. The modeling is included in a unified system with the monitoring as a presently widespread method of observation for and analysis of actual information and system development. The formalization of state processes and of dynamics of general and special parameters in the form of mathematical mappings underlies the proposed method. In a generalized form the model of system development and hierarchical control is represented as follows.

Let $F_{ij}$ be a set of control actions for $ij$-th subsystem, $F_{ijk}$ a subset corresponding to $k$-th state, $[0, T^*]$ a control time interval, and $f(i, j, t) \subseteq F_{ij}$ a control action on $ij$-th subsystem at time moment $t \in [0, T^*]$. Then the control process is described by the vector-function

$$f(i,j,t) = (f_1(i,j,t), f_2(i,j,t), ..., f_n(i,j,t)) \qquad (1)$$

in control space $\prod_{i=1}^{n} F_i$ of Cartesian product of sets of control actions on different subsystems. We consider that $f(i, j, t)$ influences uniquely on subsystem state and on the value of its efficiency criterion.

Let $s(i, j, t)$ be a process of state changes, and $w(i, j, t)$ a process of efficiency criterion changes for $ij$-th subsystem at control time interval $t \in [0, T^*]$. Then the vector-functions

$$s(i,j,t) = (s_1(i,j,t), s_2(i,j,t), ..., s_n(i,j,t)) \qquad (1')$$

and

$$w(i,j,t) = (w_1(i,j,t), w_2(i,j,t), ..., w_n(i,j,t)) \qquad (1'')$$

describe the attainable configurations that represent the efficiency of control process $f(t)$ at the moment $t \in [0, T^*]$.

The main restriction of monitoring as a method of analysis of system development is that it is in direct relationship to the quality of data gathering organization (scheme), which provides compatibility and co-dimensionality of data that represent separate components of analyzable object at different observation moments. For this reason, preceding the modeling of complex objects one should preliminarily design the structure of monitoring. In general case, monitoring should be presented as a multi-step and multi-aspect system, including both independent and information-dependent monitorings. To construct interdependent development models and to transit to the development models of the further level of hierarchy, monitorings should be combined into a system (scheme) of parallel observation for the process of development of object under investigation. Multi-step and multi-aspect monitoring provides system approach to the modeling and correlation of information, which is used for decision making on the various related control problems.

In a generalized form, a system of monitorings can be presented as a multilevel object-oriented observation system (fig.2).

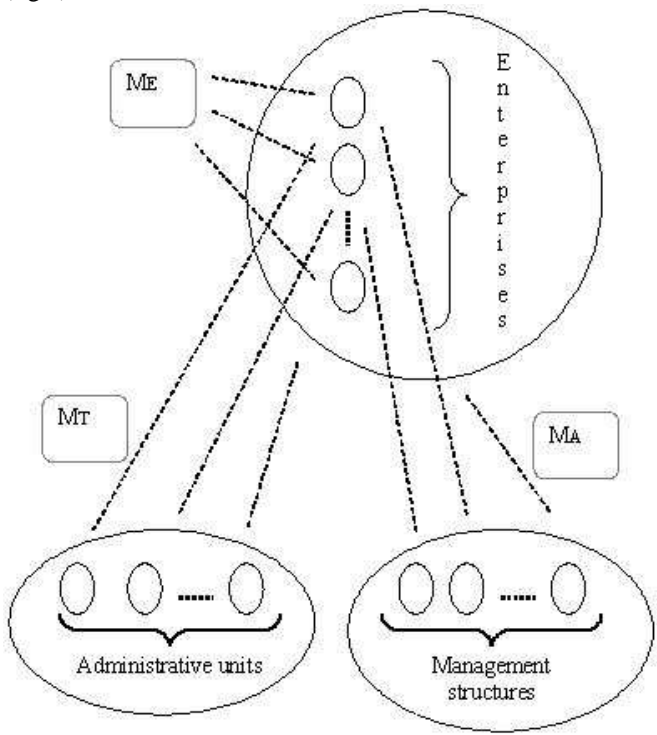

Fig. 2. Organization of monitoring systems

Figure gives an example of structural design of monitoring in the framework of long-term regional control. The data provided by M$_E$, individual monitoring of industrial and social infrastructures of region, is used by M$_T$ which is aimed at analysis of tendencies taking place at the level of administrative units of region (territories, provinces, etc). Parallel monitorings M$_A$ are intended for analysis of efficiency of control processes in various spheres, for example, ones of public welfare, ecology, etc. In order to use monitoring data (information) on the basis of a general development model enabling to add new specialized models, a high level of semantic representation of the results of dynamic analysis of information concerning the situation around a system should be provided.

So, the more comprehensive the monitoring organization, the more efficient the modeling will be.

*B. Principles of Construction of Development Model of Complex Hierarchical System in Control Loop*

The scalability, i.e. the simultaneous representation of goals and development character of various object components, is very important when constructing a model of large-scale system development. Two factors play an important role in providing the scalability. The first one is the systematic use of hierarchical principle for representation of control object, control system, and system parameters. The second one is to establish polymorphic parameters equally applicable for objects at different levels of hierarchy. Polymorphic parameters in hierarchical models enable one to turn from control at the object level to control at the level of object classes, and also from individual models to integral models of arbitrary level of generalization. In this case, the problem of modeling of complex large-scale hierarchical system development can be reduced to the analysis and interpretation of long-term dynamics of polymorphic hierarchical parameters of hierarchical object. The most important properties of the model are:

1. Object development models at each level are sufficiently autonomous. This provides a sufficient degree of decomposability and therefore flexibility large-scale models construction;
2. Modeling objects are not only separate components but also classes of components having common development goals;
3. Models of system components are turned out to be information compatible, outputs of one component can serve as the inputs for another component.

Except for scalability, another important requirement to development models of complex hierarchical objects is to use abstract concepts for qualitative description of long-term dynamics of parameters. The basic idea concerning the abstract representation of process dynamics is to use state diagrams.

Because of the monitoring information incompleteness, fuzziness and possible uncertainty the advantage of the appropriate informative method of interpretation of the processes of changes in the values of parameters should be made in order to make it intuitively clear the splitting of time series into intervals which differ from each other by the character of control actions.

The continuous time interval $[0,T^*]$ is divided into parts. On each part a process is described by a state, and interaction between parts is described by state diagram. A state of hierarchical object is defined as a situation which is characterized by a set of states of object components. Each state is a set of characteristic trajectories of parameters changes (fig. 3).

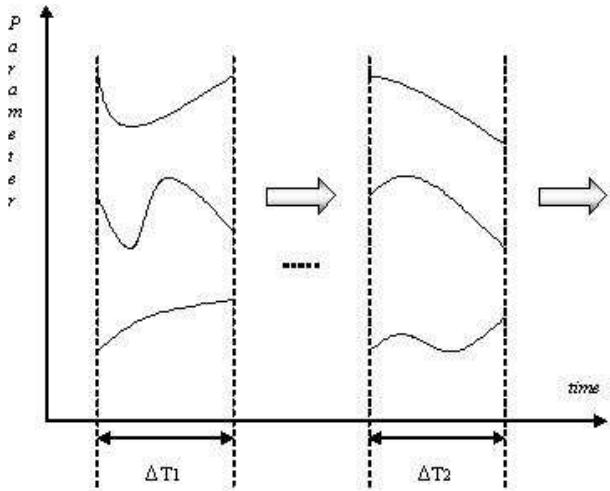

Fig. 3. Definition of state as a set of trajectories

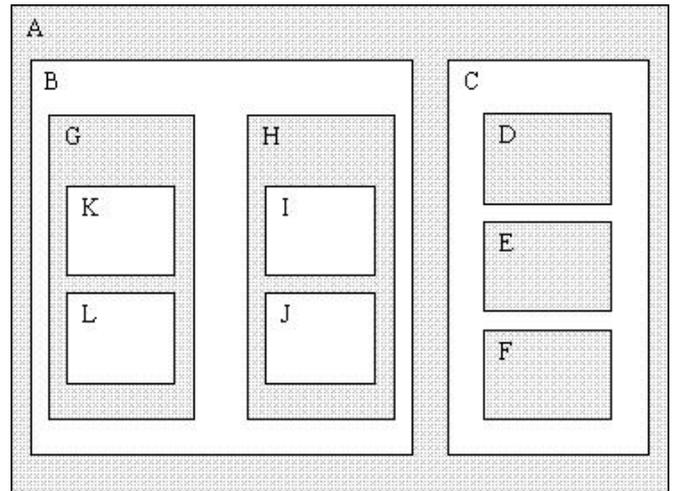

Fig. 4. Decomposition of a hierarchical state

Deformation of trajectory character on transition from one part to another formalises the state change. The state change can be used for estimation of direction, efficiency and quality of control actions. In this way we get a qualitative image of state dynamics which is essential for control goals. This helps represent the real and desirable characteristics of control object, their properties, structural and functional interrelations. The approach to the representation of system states has the following properties:

1. aggregates the parameters and, therefore, simplify the system modeling;
2. formalises the information gathering and estimation for getting the integrated and local evaluation of hierarchical system;
3. forms the basis for analysis of system dynamics and facilitates the study of a number of aspects of dynamical process in a unified way.

Fig. 4 illustrates a decomposition of hierarchical state, each component A, B, C, D, E, F, G corresponds to a set of parameters of a certain level of hierarchy. The appropriate semantic interpretation used, the hierarchical state shows how the current states of objects of different levels of hierarchy are related to each other.

To turn the language of hierarchical states into convenient and efficient tool, we construct a formalized model which is based on the universal decomposition scheme of development model of hierarchical object.

Controllable objects belonging to a hierarchical set $W$ are estimated by polymorphic parameter of hierarchical structure $I$. Each component is represented by state diagram from a set $D$ of state changes (real, desirable, and predictable) (fig. 5).

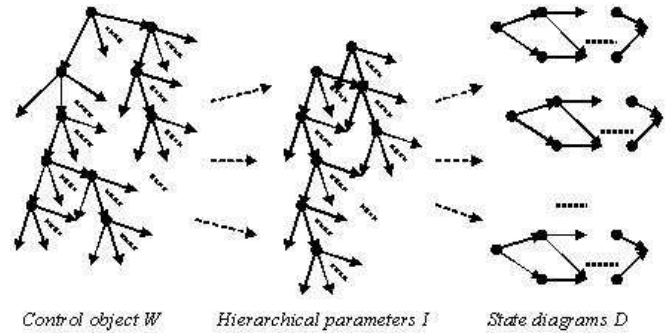

Fig. 5. Generalized structure of decomposition of development model

The diagrams from $D$ that correspond to the objects of one level of hierarchy represent sequential processes of state changes. The diagrams of objects of higher level of hierarchy represent the corresponding (parallel, taking place at the same time moment) states of higher level.

*C. Formalised Scheme of Construction of Development Model of Complex Hierarchical System with Polymorphic Parameters*

In this subsection we give a scheme for construction of development model of objects of one level of hierarchy. The scheme is a basic algorithmic step for construction of development model of complex hierarchical system with polymorphic parameters.

Let us denote the set of objects of one level of hierarchy $W'$. The scheme is divided into four stages.

The *first stage* includes the preliminary study and consists in establishing the parameters with parallel dynamics which characterize an arbitrary object from $W'$. At first stage we choose the set of parameters and form the graph representation of their parallel dynamics at a given time interval. Using a graph representation we compare the character of parameters changes of object under study. The *second stage*, the stage of dynamic parameters estimation, consists in getting the comparative dynamical characteristics of polymorphic parameters for different objects from $W'$, and in extrapolating the dynamics of parameter values for arbitrary object with simple relationships, which describe the essence of processes under study. The analysis of parameter dynamics gives answers to the following questions: whether a parameter is a function of time of any standard type, monotone increasing or decreasing, with one or several critical points, whether the function is bounded, whether it has a point of inflexion, or it can be described by a cyclic process. The basic idea of algorithm for recognition of type of the dynamic process consists in estimation of state of parameter dynamics. This includes heuristic analysis of a sequence of parameter values $X(1)$, $X(2),...$ and producing the current state process estimate $S(t) = F(S(t-1), X(t-1), X(t))$ in arbitrary time moment. The algorithm is universal and applicable for any parameter, for which values the notion of comparison is defined. A qualitative estimation of the current process of a parameter dynamics enables one to create diverse classification rules for objects from $W'$. This activity forms the *third stage*. Classification rules are given by means of matrices with logical elements. An element $(I, J)$ of matrix, where $I$ is a parameter and $J$ is a class of objects, $J \subseteq W'$, contains a logical formula which determine the current state of process of $I$-th parameter changes. The matrices of this type give rules of one-level classification. However, the most important are rules of hierarchical classification based on the eventual specification of conditions to be satisfied by objects from a class. The subclass of multilevel classification rules which along with the grouping of objects reflects the semantics of states development of these objects is of the most interest. The notions of state scale and classificator are the formal basis for construction of multilevel classification rules.

Let $K = \{k_1, k_2, ..., k_q\}$ be a set of predicates, propositions relating to the parameter values of objects set $\Omega$. The ordered set of predicates $K = \{K_1 < K_2 < ... < K_n\}$, $T_{K_i} \cap T_{K_j} = \emptyset$, where $T_{K_l}$ is a truth domain of $K_l$, $l = \overline{1, n}$, is called a one-level scale (or simply scale) if each $K_l$ defines a state $S_l$. It is assumed that the predicates and the corresponding states have the same ordering, i.e. if $K_1 < K_2 < ... < K_n$ then $S_1 < S_2 < ... < S_n$. The scale determines the values of parameters and enables us to compare the states of the objects.

We say that the scale $\{K_{i_1} < K_{i_2} < ... < K_{i_n}\}$ is a hierarchical continuation of the scale $\{K_1 < K_2 < ... < K_i < ... < K_n\}$ if the predicates $\{K_{i_1} < K_{i_2} < ... < K_{i_n}\}$ are the set of sub-predicates of $K_i$. A hierarchical system of scales is called to be a classificator of objects from $\Omega$ over the hierarchical set of parameters at time interval $\Delta$. At the *fourth stage* the classificator is used for formal description of dynamic development model of objects $W'$. Formalized scheme of dynamical system description is given in the form of canonical model of state development of objects $W'$.

A canonical model of state development of a set of objects is represented at time interval $[0, T^*]$ by state transition diagram

$$D = \{S, K, P, S_0, S^*, (\mu_i, i = \overline{1, n}), \mu_0, \mu^*\} \quad (2)$$

where

$S$ - a set of states ordered by $K$,

$S_0$, $S^*$ - initial and final states respectively,

$P$ - a set of arcs; each arc is assigned a time interval $\Delta \in [0, T^*]$ of state transition, if $(S_1, S_2) \in P$ then $S_1 < S_2$,

$\mu_1, \mu_2, ..., \mu_n$ - a sequence of objects distributions over the nodes-states of the diagram at time moments $t_1, t_2, ..., t_n$ respectively; $\mu_0$ - an initial distribution, $\mu^*$ - a final distribution (fig. 6).

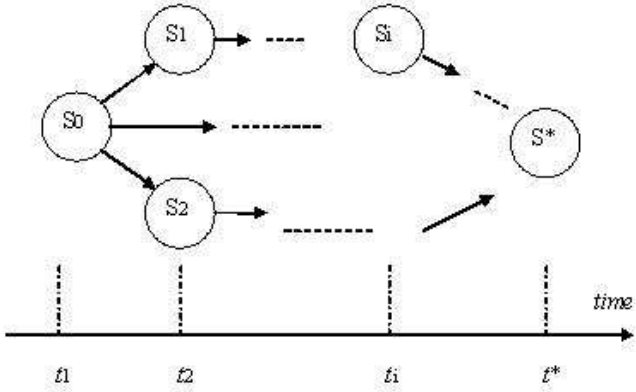

Fig. 6. A canonical model of development of objects' states

The canonical model formalizes the qualitative properties of dynamical system and represents a hypothetical model of development. A hypothesis about the character of state changes of objects set is formulated as follows. If states of the objects $W'$ at initial observation moment have (are defined by) the distribution $\mu_1$, then the dynamics of the objects states can be presented by the distributions $\mu_2,...,\mu_n$ at time moments $t_2,...,t_n$, respectively, and by the distribution $\mu^*$ at the final time moment. For example, if $n=2$, whether it is true that if all the objects $W'$ at the initial time moment are in state $S_0$, which has the lowest qualitative estimate, i.e. $\mu_0 = \{W' \Leftrightarrow S_0\}$, then at the final time moment all the objects transit to state $S^*$, which has the highest qualitative estimate, i.e. $\mu^* = \{W' \Leftrightarrow S^*\}$. The canonical model is used for comparative analysis with the real state dynamics of a set of objects. To compare, it is necessary to perform re-estimation of states of the objects at the sequential time moments $t_1, t_2, ..., t_i, ..., t_n$ in order to get the real distributions of states of the objects over canonical states of the diagram, and then to compare this distribution with the required one. This helps represent the core of system control and development problems.

The description of real development process at arbitrary time interval $[t_i, t_j]$ is based on the use of states of canonical model as objects classifier. To the set of arcs $P$ in canonical model a set $P^0$ is added. $P$ and $P^0$ are called the arcs of state development and the arcs of critical backstep of state, respectively; if $(S_1, S_2, \Delta) \in P$ then $S_1 < S_2$, otherwise, if $(S_1, S_2, \Delta) \in P^0$ then $S_2 < S_1$. $(S_1, S_2, \Delta) \in P$ means that an object from $W'$, being in state $S_1$ at time $t$, transits to state $S_2$ at time interval $t + \Delta$. Each arc $(S_1, S_2) \in P \cup P^0$ is assigned the objects counter $\eta$, which changes their state from $S_1$ to $S_2$ at time interval $[t_i, t_j]$. The counters assigned to the arcs $P$ characterize the intensity of development processes; the counters assigned to the arcs $P^0$ estimate the intensity of negative processes in object development. Consider the number of objects $N_i$ having a fixed state $S_i$ and the counter $\eta_{ij}$ assigned to $(S_i, S_j)$ as functions of time, $N_i(t)$ and $\eta_{ij}(t)$, on a observation time interval. Introduced variables enable us to get the information concerning the relation between processes of development and degradation, and the dynamics of processes (fig. 7).

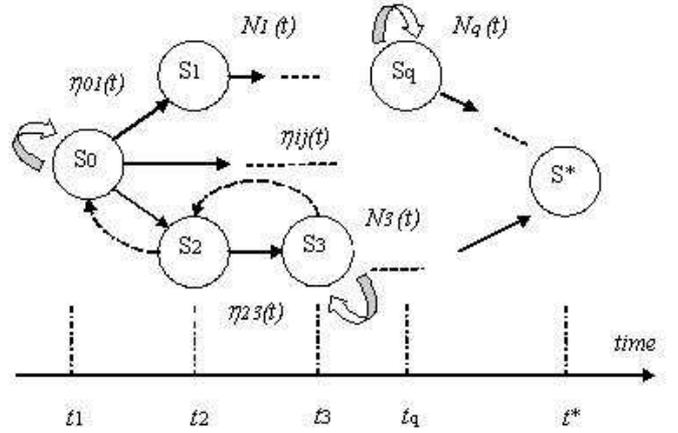

Fig. 7. State transition diagram of real development processes

This allows us to get a qualitative image of the development processes of the dynamical system under study.

The given above four stages comprise the general scheme of study of objects set of arbitrary level of hierarchy as a unified dynamical system.

IV. STATE DIAGRAMS AND DEVELOPMENT MODELS OF COMPLEX HIERARCHICAL SYSTEMS

This section is devoted to the use of state diagrams for construction of development models of complex objects. The state diagrams technique is a tool for solving a wide range of problems, estimation of control actions, comparison of control actions sets for efficiency, qualitative estimation of processes of system development, and control problem solving.

A. *Operations with State Diagrams and their Coordination*

In subsection 2.3 we proposed the analytical description of objects dynamics of one level of hierarchy. Considering the objects from neighbor levels of hierarchy, one can, in principle, create complexes of development models $\Omega = \{\Omega_1, \Omega_2, ...\}$, $\Omega_i$ are called elementary. The elementary development models enable one to analyze a number of various aspects of hierarchical object development. However, for large-scale objects the process of analysis of such models may result in a difficult problem. It is preferable to consider the models of state dynamics of multi-component systems, with each subsystem having its own local and global goals.

The state diagrams technique is universal tool for representation of dynamic development schemes for diverse control problems, not depending on their level and character. This forms the basis for coordination and consistency (concordance) of control problems. In this connection, functional generalization of several elementary models of $\Omega$ and construction of complex development models is of interest. Complex development models combine the requirements to the different sets of parameters and represent the conditions for coordination of states of the objects at different levels of hierarchy. Structural composition of state diagrams of several elementary development models provides a synthesis of complex requirements set to the dynamical characteristics of controllable object. The structural composition holds a central position in the models of hierarchical system dynamics.

Let $D = \{D_1, D_2, ..., D_i, ..., D_n\}$ be a set of diagrams to be composed, given at time intervals $\{[0, \tau_1], [0, \tau_2], ..., [0, \tau_i], ..., [0, \tau_n]\}$ respectively. Then, we say that for diagrams $D_i$ the property of consistency holds if the attainability of certain states takes place in a given (prescribed) time-event sequence.

We introduce the basic operations for state diagrams in order to give criteria for their coordination as follows.

I. Sequential-parallel composition

a) a set of diagrams $D$ forms a linear fragment, sequentially composed, if for their time intervals the following inequality holds $\tau_1 < \tau_2 < ... < \tau_i < ... < \tau_n$ (fig. 8a).

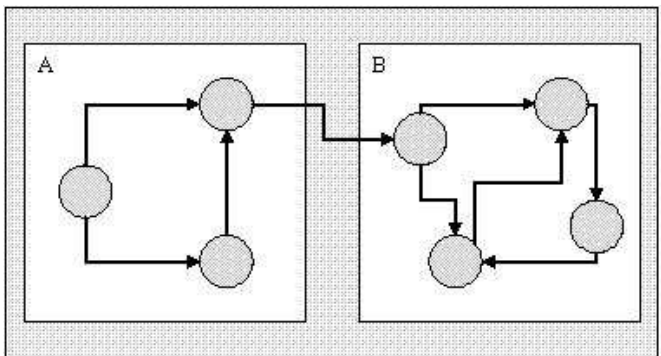

Fig. 8a. Sequential composition

b) a set of diagrams $D$ forms a parallel fragment, composed in parallel (fig. 8b), if they are defined on the same time interval.

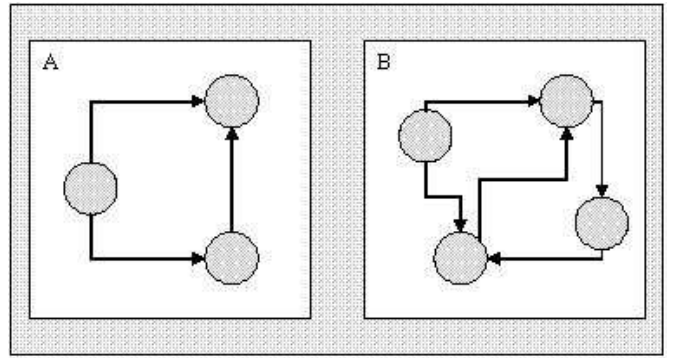

Fig. 8b. Parallel composition

II. Generalization

To give criteria for coordination of dynamical systems at neighbor levels of hierarchy we use the Cartesian product of states of diagrams of lower level of hierarchy (fig. 8c).

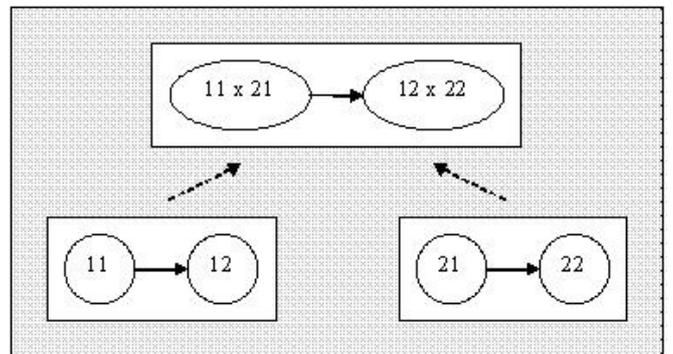

Fig. 8c. Generalization

In this case, the coordination criteria for state development of dynamical system at neighbor levels of hierarchy is realized by specifying the ordering relation on the subsets of Cartesian product of states of diagrams of lower level of hierarchy. Fig. 9 illustrates the operation of generalization. In the example, the development process of higher level is considered as a two-stage: the first stage is to complete the processes of transitions through the states $\{S_{11}, S_{12}, S_{13}\}$ and $\{S_{21}, S_{22}, S_{23}\}$ in children dynamical system; the second stage consists in the transition to the final states $S_{16}$ and $S_{26}$.

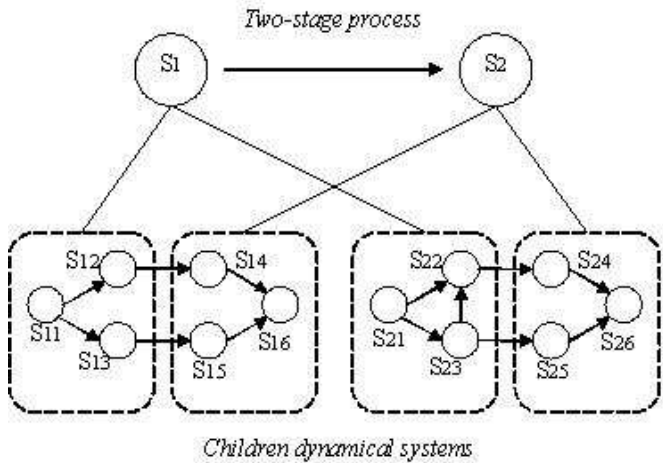

$S1 \in \{S11, S12, S13\} \times \{S21, S22, S23\}$
$S2 \in \{S14, S15, S16\} \times \{S24, S25, S26\}$
$S1 < S2$

Fig. 9. Operation of generalization for two-level hierarchy

The composition of diagrams allows one to formally represent different combinations of complex criteria sets to perform objects classification and to solve control problems. Using the consistency rules and operations with diagrams one can model diverse schemes of inter-level relations and influence of states of lower level diagrams on the processes of higher levels of hierarchy. As a result, a certain value is produced at the output of the highest level.

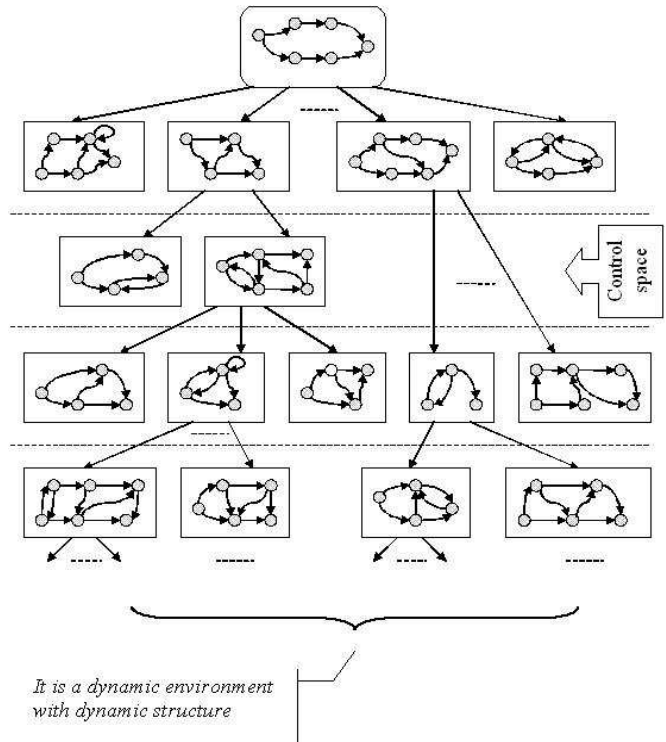

Fig. 10. A model of hierarchical network

This value is considered as a response, reaction, of the whole *hierarchical network* (fig. 10) on the values of input parameters. In accordance with ideas and the proposed notions of system dynamics, it is expedient to use the notion of *hierarchical graph (network) automaton* equally with the concept of cellular automaton.

*B. Model of Controllable Development*

Elementary and complex state diagrams enable one to construct (design) clear and graphic development models. The nodes of diagrams are states, and the arcs are intensities of objects transitions from one state to another. The ordering of states demonstrates the process of objects development. Using the modeling tools, the development model can be redefined and new states and new ordering relations can be given. The study (considering) of state diagrams in time domain (time-domain analysis) – animation of objects' flows – allows forming the time characteristics of the process of development of a set of objects under investigation (study).

The various approaches to the control processes require consideration of development models, in which the representation of controllable dynamics of hierarchical object initiated by input signals comes to the forefront. The model of controllable development is based on the following principles:

1. selecting the control actions that influence the controllable system; this is important for autonomous construction of control scenario and for flexible modification of the model to the alternative control scenarios;
2. taking into account the states that have been attained on the previous control stages (system state history); this provides a succession of multistage control scenarios;
3. comparing with the results of alternative control scenarios; this provides basic arguments upon estimating the efficiency of control scenarios.

The model of controllable development illustrates the key dynamic characteristics depending on whether or not the control actions corresponding to the current states are performed. In this sense, the model of controllable development is constructed in the form of hypothesis "what-if…"

A hypothesis is defined by state transition diagram

$$D^H = \{S, P, S_0, S^*, X\} \qquad (3)$$

where

$S$ - a set of states,

$S_0, S^*$ - an initial and final states respectively,

$X$ - a set (alphabet) of input control symbols,

$P$ - a set of arcs; $P = P_1 \cup P_2$, $P_1 \cap P_2 = \varnothing$,

$P_1$ - a subset of arcs of state transitions initiated by input symbols,

$P_2$ - a subset of arcs of state backstep in the absence of input symbols,

$X \Leftrightarrow P_1$ - a correspondence that determines for each input symbol the state transition initiated by the symbol.

To model and analyze the connections between different subsystems we introduce a mechanism of win/loss that other subsystems can obtain depending on the state of each element.

Let us denote the input alphabet $X = \{\cup X\}$, the set of arcs $P = \{\cup P_1\}$, of a set of state transition diagrams.

We define the mechanism of after-effect by splitting $P$ and $X$ into two subsets $(Z, U)$ and $(X^Z, X^U)$, respectively. The arcs of $Z$ are called isolated, and arcs of $U$ are called coupled. According to this, symbols of $X^Z$ are called individual (or special-purpose), and symbols of $X^U$ are called general (or general-purpose).

In order to define a mechanism for coupled arcs we introduce the parent-arcs as a Cartesian product of child-arcs for state transition diagrams of subsystems of neighbor levels of hierarchy. The isolated arcs $Z$ represent the state transitions initiated by individual input symbols $X^Z$; this kind of symbols do not influence on the state transitions of other subsystems. The coupled arcs $U$ represent the state transitions initiated by general input symbols $X^U$; this kind of symbols initiate the state transition on the parent-arc, which means, as a consequence, the state transitions on the corresponding child-arcs. And conversely, state transitions on all or several child-arcs can initiate a state transition on the parent-arc of subsystem of higher level of hierarchy.

A model of scenario controlling the development of control object is a 5-tuple

$$\{\Omega, I, M, C, V\} \quad (4)$$

where

$\Omega$ - a system of state transition diagrams; they represent the programs of state changes for each subsystem,

$I$ - the hierarchical structure,

$M : I \to \Omega$ - a functional that assigns a hierarchical number to each diagram of $\Omega$,

$C$ - time diagram for symbols $X$; it determines the sequential-parallel process of input symbols entering,

$V$ - a scheme of after-effect of state transitions.

To give the time diagram $C$ of input control symbols entering one can use various ways, including the estimation rules of each current state of system.

The trajectory of attainable states represents general and local goals solved by scenario on arbitrary time interval. The study (investigation) of the basic properties of scenario is reduced to the analysis of trajectory of attainable states and its comparison with the expected or predicted effect. Some of the examples are:

1. Completeness of scenario; this means the transition of all subsystems to the final states of the corresponding state transition diagrams;
2. Redundancy of scenario; this means that the input symbols (signals) of different types, individual and general, enters the input of subsystem;
3. Omitted possibilities of scenario; this is exhibited by transition frequency on the arcs representing the backstep of the attained state;
4. Complexness of scenario in problem solving is estimated by transition frequency on the coupled arcs.

V. ARCHITECTURAL MODEL OF INFORMATION SYSTEM FOR SIMULATION AND ANALYSIS OF COMPLEX SYSTEMS

In this section we propose a general structure of information system focused on the support of simulation and analysis of dynamic processes and scenario control efficiency in complex systems, based on the developed original mathematical tools.

*A. Description and requirements to computer system*

One of the most important elements providing adequate representation and modeling of dynamics of complex systems, and also the analysis of their development is knowledge of values of various parameters of system and tendencies of their change, both in a mode of absence of external influences and in arbitrary control loop. Systems of monitoring allow users carry out observing for the current values of parameters and for the actual information on the character of system development. This, in turn, allows one to estimate conditions and to predict possible (probable) events in a system and consequences following from them which can be caused by changes in values of parameters.

Forecasting of events can be based on logic of the retrospective analysis, the essence of which is the following. When forecasting events, the parameters of system are continuously measured. If there was some event in a system and for some time before the event a parameter has sharply changed, or there was a gradual change of values of parameter up to some critical, then such anomaly is related with this event. The dependences of such a kind confirmed repeatedly, i.e. becoming steady, are used for estimation and forecasting of possible future events in system. Actually, knowledge and experience obtained in the past and expert knowledge are used.

The corresponding information system should provide:
- Identification and registration of the information on the occurred events and on the current

- situation;
- Information storage and maintaining;
- Information usage by gathering, aggregations, classifications, processing, and delivery of requested necessary information.

Along with the information functions the possibilities of modeling and forecasting of events succession at the realization of alternative control strategies should be stipulated.

In this context, as a model solving the aforementioned tasks the model of directive planning is considered.

*B. Architectural model of directive planning system*

The purpose of directive planning is construction of sound and efficient scenarios of development of objects under investigation. The essence of the problem analysis that arises in this connection is:

- To reveal, establish, and demonstrate core points of a set of interconnected control problems;
- To present different approaches to the solution of control problems; to compare the approaches for efficiency;
- To organize delivery of analytical documentation with the conclusions confirmed with the graphic or statistical information.

It is reasonable (expedient) to carry out all of the tasks of directive planning within the framework of computer system that offer users all the necessary information resources and means of analysis.

A generalized structure of directive planning system (fig. 11) includes: user interface, parameters library, builder of canonical model of development, monitoring databases, interpreter of monitoring database, and model of controllable development.

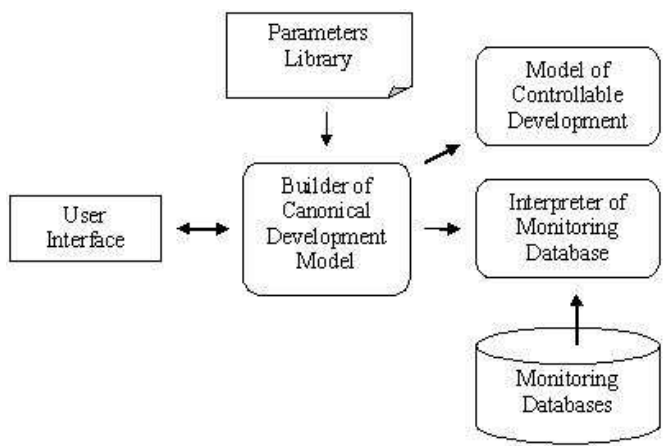

Fig. 11. A generalized structure of directive planning system

As the system should support continuous observing process for a number of parameters, including problem-oriented ones, it contains the library of hierarchical blocks of parameters which is extendable and editable. For example, applied to problems of regional development, the library can include blocks of parameters of social status and living standard, ecology, level of development of economy, etc. The builder of canonical model of development represents a specialized system of entering of state diagrams as input information. The state diagrams tool enables clear and precise formalization of states, inherent for one process and uncharacteristic for others. Therefore, it can be used for representation of regularities and typical models of states development.

Directive planning considered in control loop assumes a high level of informatization and operative connection with monitoring database.

The interpreter of monitoring database and the model of controllable development are the basic components in the presented information system. The interpreter of monitoring database functions according to the composition of canonical models of development specified by the user and generates the description of real (actual) multilevel dynamics of hierarchical object (fig. 12).

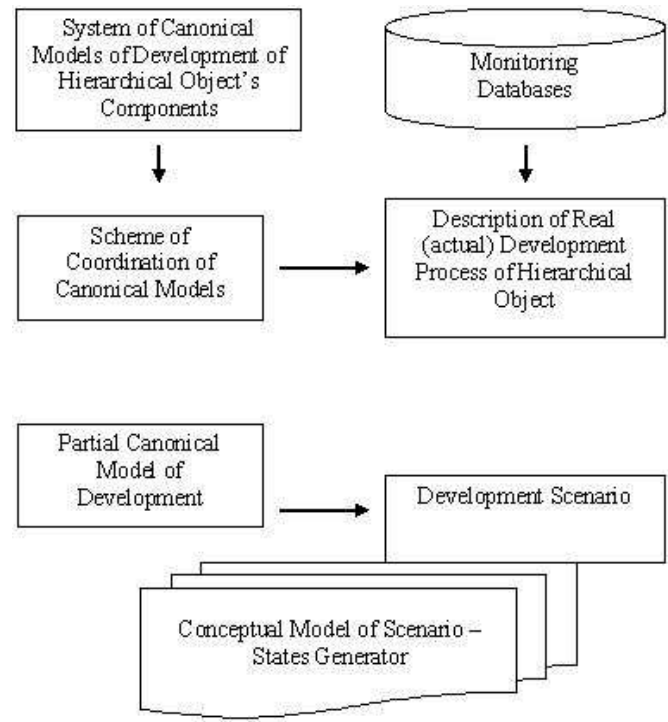

Fig. 12. Multilevel dynamics of hierarchical object

The model of controllable development is constructed as expert subsystem for assessment (estimation) of control scenarios defined by the user.

The development scenario being estimated plays a role of an inference system and is considered as generator of consecutive states of object under investigation. The rules of states generator are represented in the format of tree-like decomposition of global goal on the sub-goals; to each terminal (final) node an elementary rule is assigned. The

format of an elementary rule IF-THEN-ELSE (fig. 13) is presented by 7-tuple

$$(w_i, S_{ij}, S_{ik}, S_{il}, P_{ik}, R_{ik}, t_{ij}) \qquad (5)$$

determining a control action $P_{ik}$ which should be undertaken to transfer a component $w_i$ from the state $S_{ij}$ to the state $S_{ik}$ in time $t_{ij}$ with the resources expenditure $R_{ik}$ while not admitting the backstep to the state $S_{il}$.

The rules of state transformations represent a convenient way for construction of control scenarios as they
- Allow one to easily realize iterative process of creation and modification of control scenarios;
- Admit the efficient realization by means of executive procedure;
- Possess the sufficient expressiveness of the specification of control processes.

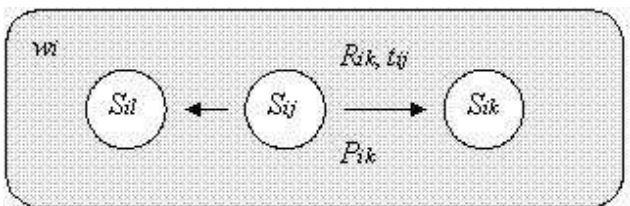

Fig. 13. A format of elementary rule If-Then-Else

The model of controllable development is used for checking a hypothesis about the efficiency (effectiveness) of the scenario being estimated. The criterion for an estimation of the scenario is given in the form of "partial" or "incomplete" state development diagram determining (specifying) support states which should be achieved with the specified restrictions on the time and resources. The construction of scenario can be divided into several stages; in each concrete case the stages can have more detailed character:
- Analysis of initial state of an object and possible trends of state changes;
- Determination of a spectrum of states of the object in the near future;
- Formulation of probable hypotheses of the object's transition tendencies from these states to the subsequent ones;
- Analysis and establishing of desirable end result – final state of the object.

The special case of "partial" or "incomplete" state development diagram is the pair of states: initial state and desirable final state.

In this case, the expert subsystem should
- *either* confirm a hypothesis that the model of development controlled on the basis of analyzable scenario meets (satisfies) the given criteria or requirements, and supplement the input diagram with the specifying intermediate states
- *or* refute the hypothesis and generate (produce) computer forecast (prediction) in the form of alternative state development diagram.

Both during the interpretation of monitoring data and in the course of scenario development, a few (several, some) loops of modeling are organized. The loops are joined (united, connected, associated) by the goals in view of the concept of representation of object as a hierarchical system, and of the corresponding operations of state diagrams compositions. The State Generator enables:
- to study effects of integrated and multi-aspect control regarding different subsystems of complex object;
- to divide the control process into stages;
- to perform decomposition schemes of prediction (forecasting) in which each subsequent model is an integrated or detailed elaboration of the previous;
- to construct and analyze the interconnected aggregated and detailed models of development parameters.

The generalized scheme of directive planning is presented in fig. 14. The scheme includes the stage of retrospective analysis and that of construction of directive model of development.

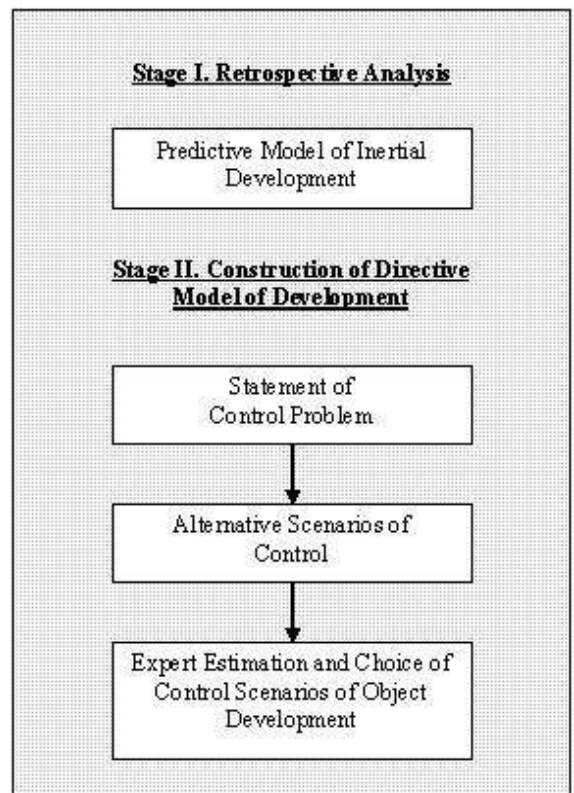

Fig. 14. A generalized scheme of directive planning

Construction of retrospective models is intended for performing tests and selection of alternative approaches to the strategic development of objects under study. The retrospective analysis, based on the usage of the *Interpreter of Monitoring Database*, consists in carrying out calculations and the subsequent estimation (evaluation) of the results for the last period. The main advantage of the analysis is the possibility of comparing actual (real) and estimated data. The results of the retrospective analysis reflect the most important regularities and trends of the previous period.

In the course of retrospective analysis the objects of investigation should be selected, the dynamics of the chosen parameters studied, and the state diagram constructed, which interprets the monitoring data.

Thus the diagnostic analysis of objects' state development, which is characteristic for pre-crisis and crisis processes, is implemented. By empirical analysis, experimenting, and selecting different system of parameters of object development the diagrams which the most expressively depict "negative" (positive) trends should be found. These diagrams serve as a means of formal representation of the current control problems and answer the question "what will be if no control actions should be undertaken". The results of retrospective analysis help put forward the goals and problems of development and to form possible alternatives of controllable development for the perspective period.

The aim of the next stage is the construction of directive model of development, analysis of controllable processes of complex objects development, and obtaining (deriving) answers to the question "what should be undertaken to achieve the required goals". At this stage an initial state and directive expected (final) state should be described, and the space of intermediate states should be constructed. Then, the conceptual model of controlling scenario in the form of *States Generator* should be defined.

Using the toolkit of dynamic expert systems, the information environment can be adapted to the current range of problems with minimal costs.

*C. Generalized structure of computer simulation system*

A model of construction of controlling scenarios in the directive planning environment provides a wide variety of possibilities for analysis of different approaches to the solution of diverse control problems, and comparison of the approaches for efficiency. The model can also provide unification of a system of parameters, goals and control object development conditions for different analyzable control action sets.

A generalized structure of computer system, based on the usage of the modeling system of controlling scenarios, for analysis and selection of efficient control actions, is presented in fig. 15.

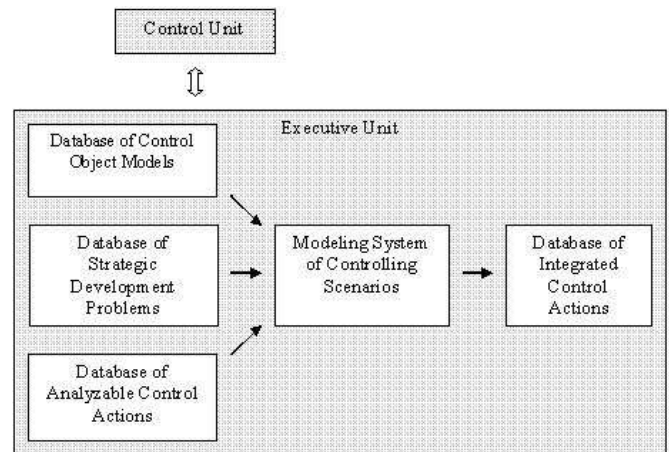

Fig. 15. A generalized structure of computer system intended for analysis and selection of efficient control actions

The model of controlling scenarios together with the database of models of control objects, the database of strategic development of these objects, and the database of analyzable control actions provides a universal environment for formal representation of goals, control and development problems, time and resource characteristics, etc.

The organization of modeling of controlling scenarios in the directive planning environment gives side benefits because of an opportunity of extension of set of criteria of control actions efficiency and their selection regarding the long-term strategic development goals of controllable object and division of development process and goals into stages and sub-goals, respectively. It creates unique conditions for analysis and comparison of alternative control actions having common initial premises and focused on the achievement of the same goals.

Consideration of control actions in the context of achievement of the desirable goals (attainable states) of hierarchical object development gives one more advantage of usage of the model of controlling scenarios. It consists in an opportunity of analysis of corporate (global) efficiency of control actions focused on the achievement of goals at different levels of hierarchy. Thus, there is an opportunity of transition from the stage of comparison of separate control actions to the stage of formal synthesis of the integrated control actions for hierarchical object as a whole and their further comparison.

The control unit realizes the functions of organization of expert estimation and comparison of control action sets, knowledge base support, and decision making. The functional subsystems of Executive Unit reproduce the basic stages of control actions selection, including the stage of synthesis and initiation of problem and control domain models.

The structure provides the fullest means for simulation and analysis of control actions.

## VI. Conclusion

We have presented the models, analysis methods, and the structure of computer information system, which are the basis for design and construction of applied systems for modeling, analysis, control, and prediction of development processes of complex dynamical systems with the use of models of controllable development of hierarchical systems. The technique presented can also be used as a technology for construction of information systems for simulation analysis of development strategies and control scenarios of complex objects, and has been applied in several information systems and decision support systems.

We presented both general and special theory. The former concerns formalization of basic concepts and techniques for schematic representation and modeling of discrete hierarchical dynamic process; the latter one specializes the formalism to modeling the coordinating scenario-type control schemes. The method allows one to model inertial system dynamics that determines the current state consequences, and to demonstrate future state dynamics of system in arbitrary scenario-type control loop.

The technique is especially powerful when applied in information-rich environments. The information can be simultaneously aggregated in a few ways: by hierarchical structure of processes and states embedding, by parallel representation of dynamical characteristics of several processes within the framework of one state, and by dividing the observation time interval in relation to the events associated with the changes in system dynamics and tendencies. The proposed models and technique are universal and at the same time it is problem-oriented in relation to the rationality, consistency and coordination of control actions; it can be equally used for diverse kinds of systems such as technical systems, organizational systems, socio-economic systems, systems of strategic planning and long-term forecasting systems, and decision support systems.

We suppose that the theoretical and computer models presented can serve as a tool for designing and modeling of complex dynamic systems with control and for designing automated information systems for analysis, simulation, and forecast of development of complex systems.

## References

[1] D. Harel, "Statecharts: A visual formalism for complex systems", *Science of Computer Programming*, vol. 8, pp. 231-274, 1987.

[2] M. P. E. Heimdahl, N. G. Leveson, "Completeness and consistency in hierarchical state-based requirements", *IEEE Trans. Software Engineering*, vol. 22, no. 6, pp. 363-376, 1996.

[3] A. G. Bagdasaryan, "A model of automated information system for solving control problems of large-scale systems", *Control Sciences*, no. 6, pp. 65-68, 2005.

[4] A. G. Bagdasaryan, "An approach to the coordination of hierarchical scenarios", in *Proc. 1st Int. Conf. Cognitive Analysis and Situation Control*, Moscow, 2001, pp. 72-73.

[5] A. G. Bagdasaryan, T. R. Ohanyan, "A model of data organization and analysis in large-scale information systems", in *Proc. 6th Int. Conf. Cognitive Analysis and Situation Control*, Moscow, 2006, pp. 374-377.

[6] A. G. Bagdasaryan, "A general structure of information expert system for simulation and analysis of complex hierarchical systems in control loop", *Control of Large-scale Systems: Sbornik*, to be published

[7] V. Kulba, D. Kononov, S. Kosyachenko, "Scenario calculus as methodology for analysis of complex systems", *Proceedings of the Institute of Control Sciences,* vol. 9, 2000.

[8] A. G. Bagdasaryan, *Discrete Modeling and Analysis of Complex Dynamic Systems in the Control Mode*. Moscow: ICS RAS, 2005

[9] T. Toffoli, N. H. Margolus, *Machines of Cellular Automata.* Moscow: Mir, 1991.

[10] M. Mesarovic, D. Macko, Y. Takahara, *Theory of Hierarchical Multilevel Systems*. N.Y.: Academic Press, 1970.

[11] M. Mesarovic, Y. Takahara, *General Systems Theory: Mathematical Foundations*. N.Y.: Academic Press, 1975.

[12] V. Latora, M. Marchiori, *The Architecture of Complex Systems*. Santa Fe Institute for Studies of Complexity, Oxford University Press, 2003.

[13] Y. Bar-Yam, *Dynamics of Complex Systems*. Reading, MA: Addison-Wesley, 1997.

[14] B. Edmonds, "Complexity and scientific modeling", *Foundations of Science*, vol. 5, pp. 379-390, 2000.

[15] M. E. J. Newman, "The structure and function of complex networks", *SIAM Review*, vol. 45, pp. 167-256, 2003.

[16] M. Wooldridge, *An Introduction to Multi-agent Systems.* Chichester: John Wiley and Sons, 2002.

[17] F. Harary, R. Z. Norman, D. Cartwright, *Structural Models*. N.Y.: Wiley, 1965.